\documentclass[12pt]{article}

\usepackage{authblk}
\usepackage{natbib}
\usepackage{booktabs, graphics, amsmath, amssymb}
\usepackage{subfigure}
\usepackage[figuresright]{rotating}

\usepackage{amsmath, amsthm, amssymb}
\usepackage{url}
\usepackage{tikz}
\usetikzlibrary{bayesnet,fit,positioning,shapes,arrows}

\newcommand{\bm}[1]{{\boldsymbol {#1}}}

\title{Where geography lives? \\ A projection approach for spatial confounding}
\author[1]{Renato Martins Assun\c{c}\~{a}o}
\author[2]{Marcos Oliveira Prates\thanks{Corresponding author: marcosop@est.ufmg.br}}
\author[3]{Erica Castilho Rodrigues}
\affil[1]{Departmento de Ci\^{e}ncia da Computa\c{c}\~{a}o, Universidade Federal de Minas Gerais}
\affil[2]{Departmento de Estat\'{\i}stica, Universidade Federal de Minas Gerais}
\affil[3]{Departmento de Estat\'{\i}stica, Universidade Federal de Ouro Preto}
\begin{document}

\maketitle

\begin{abstract} % abstract
Spatial confounding between the spatial random effects and fixed effects covariates has been 
recently discovered and showed that it may bring misleading interpretation
to the model results. Solutions to alleviate this problem are based on decomposing 
the spatial random effect and fitting a restricted spatial regression. 
In this paper, we propose a different approach: a transformation of the geographic space 
to ensure that the unobserved spatial random effect added to the regression 
is orthogonal to the fixed effects covariates. Our approach, named SPOCK,  
has the additional benefit 
of providing a fast and simple computational method to estimate the parameters. 
Furthermore, it does not constrain the distribution class 
assumed for the spatial error term. A simulation study and a real data analysis are presented
to better understand the advantages of the new method in comparison with the existing ones.
\end{abstract}

\section{Introduction}

Spatial generalized linear mixed models (SGLMM) for areal data analysis have become
a common tool for data analysis in recent years with the availability of spatially referenced data sets.
\citet{besag:york:mollie} 
introduced a hierarchical modeling adding random spatial effects to a
generalized linear regression model. The spatial dependence is
captured by a latent Gaussian Markov Random Field (GMRF). 
One important advantage of this GMRF approach is to induce a sparse 
precision matrix that allows for intuitive conditional interpretation 
and fast Bayesian computation \citep{rue:book}. In the past 20
years, \citet{besag:york:mollie} model (hereafter ICAR) has become the most popular areal model for its flexibility and due to its implementation in the WinBUGS software \citep{bugs}, which is freely available.

Most spatial data are observational rather than experimental and we commonly observe
correlation or multicollinearity between the covariates, 
also called explanatory or fixed-effect variables. 
As a consequence of this multicollinearity, in non-spatial regression problems,
the estimated coefficients are affected by the presence of the other 
covariates \citep{MostTukey:1997}. This is
called confounding and it can lead to implausible estimates.
It also affects the variance of the covariates' coefficients estimators which are
inflated with respect to what one would have in case the covariates
were orthogonal to each other.

In the spatial context, \citet{Clayton} and \citet{Reich} identified 
the existence of confounding between the fixed and random effects in SGLMM.
In their work, \citet{Reich} show that explanatory variables having a spatial pattern
may confound with the spatial random effects, resulting in fixed effects estimates that are
counterintuitive. Thus, they propose an alternative
model, called hereafter RHZ model, to alleviate this confounding problem. The RHZ model
projects the spatial effects into the orthogonal space spanned
by the explanatory variables. 

Another well known shortcoming of SGLMM is the computational burden when dealing with high
dimensional latent effects. Recently, \citet{Hughes} introduce an alternative
model (hereafter, HH model) that alleviates spatial confounding and, at the same time,
requires less computational effort. They also consider an orthogonal projection of the spatial 
effects in a way that takes into account the explanatory variables and the spatial structure.  
Properties of the HH model was further studied by \citet{Murakami2015}.

The main idea behind the \citet{Reich} and \citet{Hughes} methods is to project the 
unobserved spatial effects vector in the 
linear space orthogonal to the explanatory variables. As a consequence, they end up with a 
precision matrix that is far from sparse losing one of the main advantages of the Markov 
random fields. Another drawback from RHZ and HH methods is that they only work with 
spatial effects restricted to follow the ICAR model. Therefore,they do not allow for more 
flexible spatial structures such as the proper CAR model 
\citep{Besag:1974,Leroux,Rodrigues:Assuncao:2012}. 

From a geostatistical perspective, \citet{Paciorek:2010}  
dealt with the effects of confounding between the spatial random effects and 
possible explanatory variables. He showed that it can lead to bias in
the parameter estimation. \citet{Hanks:2015} also studied the effects of 
spatial confounding in the continuous support situation. They proposed a
posterior predictive approach to correct a bias on the credibility intervals 
produced by RHZ. 

In this paper, we adopt a different approach to deal with spatial confounding in lattice data. 
Rather than removing the explanatory variables effect from the spatial random effects, 
we alter the spatial pattern of the map by purging the covariates from the geography. 
We do this by projecting the spatial coordinates of the areas into the orthogonal space
of the covariates and hence producing a new set of geographical coordinates. These new coordinates
induce a different neighborhood structure for the areas that are used to 
define a different precision matrix. The transformed geography retains the spatial 
neighborhood that is orthogonal to the covariates. One important consequence of our 
approach is that we maintain the sparsity of the spatial effects precision matrix 
allowing for very fast Bayesian computation while controlling for spatial confounding. 

Our new approach is called SPOCK, an acronym for \textit{SPatial Orthogonal Centroid
``K''orrection}. SPOCK led us to understand better the role of spatial confounding and its effect 
on parameter estimation. It gives a more clear understanding of the conceptual 
differences between the parameters in models that do and do not alleviate spatial confounding.
In a striking example, \citet{Reich} showed that spatial confounding may
provide counter intuitive results in some situations. Sign and relevance of covariates 
can change drastically after one adds a spatial random effect in a lattice-data regression model. 
Some important questions arise. When does that occur? Is it 
always necessary to correct for spatial confounding? If not, when to
do so? There is an on-going discussion among the researchers about the 
need for spatial confounding correction and the meaning of the resulting 
fixed-effect parameters \citep{Paciorek:2010,Hanks:2015}. 
With our model as a framework, in the simulation study of Section~\ref{s:sim}, 
we revisit these questions and try to better understand what are the consequences of correcting for
spatial confounding and when it is adequate to do so. 

The contributions of this paper are the following: 
\begin{itemize}
    \item a new conceptual approach, SPOCK, to handle spatial confounding;
    \item because our approach retains the Markovian property, it is extremely efficient 
    for Bayesian computation even in high dimension problems generated by 
maps with very large number of areas;
\item in contrast with the present alternatives, SPOCK has 
no restriction to the type of spatial 
structure for the random effects; 
\item SPOCK is very simple to implement and it can be run in any 
Bayesian spatial software such as WinBUGS and R-INLA \citep{rue:inla}.
\end{itemize}

The paper proceeds as follows. In Section~\ref{s:existing} we review the traditional
SGLMM model and present a summary of the RHZ and HH methods. Section~\ref{s:method}
introduces SPOCK and its properties are discussed. In Section~\ref{s:sim},
a simulation study is performed to compare the proposed method against RHZ and HH methods
in terms of precision of estimates and time. It also provides insights on when to
correct for spatial confounding and what are the consequences of it.
In Section~\ref{s:real}, we revisit the Slovenia data \citep{Zadnik} to illustrate the
conclusion obtained with the three models and their computational efficiency.
Finally, the paper concludes with a discussion in Section~\ref{s:conc}.

\section{Existing Methods}
\label{s:existing}

% introduzir o modelo de Besag e explicar os modelos do Reich and Hughes.
\subsection{SGLMM}
\label{s:sglmm}

Spatial Generalized Linear Mixed Models (SGLMM) is a wide class of models
that accommodates spatial dependence through a random effect term. Let $Y_i$
be the observation of an area $i$, $i = 1,\ldots,n$ with distribution given by
\begin{eqnarray} \nonumber
 Y_i \sim \pi(y|\mu_i,\bm{\delta}, \bm{X}_i, \beta) \\ \nonumber
 g(\mu_i) = \bm{X}_i \bm{\beta} + \theta_i,
\end{eqnarray}
where $\bm{\beta}$ is the fixed effects coefficients vector and
$\bm{X}$ is a $n \times q$ full-rank design matrix with the covariates.
Typically, $\bm{X}$ includes a first column of constant values equal to 1. 
We let $\bm{X}_i$ be its $i$-th row,
$g$ is an appropriate link function, and $\mu_i= E(Y_i|\bm{X}_i)$.
The vector of hyperparameters of the distribution is $\bm{\delta}$. Finally,
$\bm{\theta}$ represents a vector with spatially
structured random variables capturing the spatial patterns shared by the
areas in study.

The most simple instance of this model is the Gaussian model where
\begin{eqnarray} \nonumber
 Y_i|\mu_i,\tau_{\epsilon} &\sim& N(\mu_i, \tau_{\epsilon} ) \\
 \mu_i &=& \bm{X}_i \bm{\beta} + \theta_i.
 \label{eq:sglmm}
\end{eqnarray}
% where $\bm{\delta}^\top = (\tau_{\epsilon}, \bm{\beta}^\top, \tau_{\theta})$.

Traditionally the spatial random effect $\bm{\theta}^\top = (\theta_1,\ldots,\theta_n)$
is defined as an intrinsic conditional autoregressive (ICAR) models introduced
by \citet{besag:york:mollie}. The prior specification of the ICAR model is given by 
\begin{equation}
\pi(\bm{\theta} |\tau_{\theta}) \propto \tau_{\theta}^{r(\bm{Q})/2} \exp{\left(- \frac{\tau_{\theta}}{2} \bm{\theta}^\top \bm{Q} \bm{\theta} \right)}
\label{eq:icar}
\end{equation}
where $\bm{Q}$ is the precision matrix, $r(\bm{Q})$ is the rank of the $\bm{Q}$ matrix 
\citep{LavineHodges:2012} and $\tau_{\theta}$ is the spatial precision. 
The matrix $\bm{Q}$ is associated with the geographical structure represented by a neighborhood 
graph $\mathcal{G}$. In this graph, each area is a node located on its spatial centroid and neighboring areas 
are connected by edges. A pair of areas $(i,j)$ that are not neighbors has $\bm{Q}_{ij}=\bm{Q}_{ji} = 0$. 
\citet{rue:book} showed that the non-zero pattern in $\bm{Q}$ defines the
conditional dependence structure of the graph under study. In other words, with the
underlying geographical neighborhood in analysis, it is possible to define the non-zero
pattern of $\bm{Q}$. Moreover, the
precision matrix $\bm{Q}$ is commonly sparse and this can be computationally explored 
to substantially improve speed in model fitting.

\subsection{Non-Confounding SGLMM}
\label{s:ncsglmm}

\citet{Clayton} introduced the concept of spatial confounding between the
fixed effects estimates and the spatial random
effects in SGLMM. However, only recently, \citet{Reich} revisited the problem
motivated by a striking case study where the credibility interval for a certain  
fixed effect coefficient changes drastically after introducing the spatial 
random effects. 
They proposed an alternative method to alleviate this confounding problem. In general,
the idea proposed is to include in the model a random effect that belong to the
orthogonal space of the fixed effects predictors. 
A new $\mathbb{R}^ n$ basis can be used to re-express 
\begin{equation}
\bm{\theta} = \bm{\theta}^X +  \bm{\theta}^{\perp} = 
\bm{K} \bm{\theta}_1 + \bm{L} \bm{\theta}_2 
\label{eq:thetaRHZ}
\end{equation}
where $\bm{K}$ is a $n \times q$ matrix that has the same span as $\bm{X}$, $\bm{L}$ is a $n \times (n-q)$ matrix whose columns 
lies in the orthogonal space of $\bm{X}$, and $\bm{\theta}_1$ and $\bm{\theta}_2$ are vectors with dimensions $q$ and $n-q$, respectively. 
Equation~\eqref{eq:sglmm} can be represented as
\begin{eqnarray} \nonumber
 Y_i|\mu_i,\tau_{\epsilon} &\sim& N(\mu_i, \tau_{\epsilon}) \\ \nonumber
 \mu_i &=& \bm{X}_i \bm{\beta} + \bm{K}_i \theta_{i1} + \bm{L}_{i} \theta_{i2},
 \label{eq:sglmm_rhz}
\end{eqnarray}
where $\bm{K}_i$ and $\bm{L}_i$ are the $i$-th rows of $\bm{K}$ and $\bm{L}$, respectively.  
The vector $\bm{\theta} $ follows the ICAR distribution in \eqref{eq:icar}.
Using this parametrization, it
was shown that $\bm{K}$ causes a confounding in the estimates
of $\bm{\beta}$. \citet{Reich} suggested the removal of the $\bm{K}$
component leading to the RHZ model:
\begin{eqnarray} \nonumber
 Y_i|\mu_i,\tau_{\epsilon} &\sim& N(\mu_i, \tau_{\epsilon} ) \\ \nonumber
 \mu_i &=& \bm{X}_i \bm{\beta} + \bm{L}_{i} \theta_{i2} \\
 \bm{\theta}_{2} &\sim& N_{n-q} \left( 0,\bm{L}^\top \bm{Q} \bm{L} \right).
 \label{eq:rhz}
\end{eqnarray}

Although it solves the confounding between the spatial random effects
and the estimates of the fixed effects, \citet{Hughes} noticed that
this correction is computationally inefficient. The reason is that the new precision
matrix generated by Eq~\eqref{eq:rhz} is not sparse and has dimension $n-q \approx n$.
To reduce the computational demand of the RHZ model
they proposed a sparse alternative model. This new model uses the so-called
Moran operator $\bm{P}^\bot \bm{A} \bm{P}^\bot$ where 
$\bm{A}$ is the graph zero/one adjacency matrix and 
\begin{equation}
\bm{P}^\bot = \bm{I} - \bm{X}(\bm{X}^\top \bm{X})^{-1} \bm{X}^\top
\label{eq:ortogonalP}
\end{equation}
is the projection matrix into the orthogonal space of the span of $\bm{X}$. They showed that
the Moran operator retains the spatial patterns of the data and it is
only necessary to select the $h  \ll n$ higher eigenvalues of the spectrum of the Moran
operator. In this way, they were able to reduce the dimension of the random effect
maintaining the spatial information necessary in model estimation. The
HH model is defined as:
\begin{eqnarray} \nonumber
 Y_i|\mu_i,\tau_{\epsilon} &\sim& N(\mu_i, \tau_{\epsilon} \bm{I}) \\ \nonumber
 \mu_i &=& \bm{X}_i \bm{\beta} + \bm{M}_{i} \theta_{i2}\\ \nonumber
 \bm{\theta}_{2} &\sim& N_{h} \left( 0,\bm{M}^\top  \bm{Q} \bm{M} \right),
%  \label{eq:jm}
\end{eqnarray}
where $\bm{M}$ contains the first $h$ eingenvectors of the Moran operator.

\section{Purging the covariates from the geography}
\label{s:method}
% introduzir o nosso modelo. Aspectos matemeticos, etc!

Although \citet{Reich} and \citet{Hughes} were successful in alleviating the
confounding and in reducing the dimension of the spatial random effects as
presented in Section~\ref{s:ncsglmm}, their approaches have two main
drawbacks: 1) the RHZ and HH models do not accept parameters in the
precision matrix $\bm{Q}$ \citep[e.g.,][]{Besag:1974,Leroux,Rodrigues:Assuncao:2012};
2) both lost the original SGLMM model sparsity of the precision matrix and hence do not take 
advantage of the Markov property in the Bayesian calculations. 
In order to propose a fast SGLMM that alleviates
the confounding problem, we introduce SPOCK, a novel approach to the problem
capable of maintaining the Markov properties of the precision matrix
as well as allowing for unknown parameters in the 
matrix $\bm{Q}$. SPOCK not only produces a fast alternative to RHZ and HH  
but, more importantly, it also represents a new and rather different conceptual perspective 
on how to deal with the problem. We will show that this new way of seeing the spatial 
confounding can help in clarifying the discussion about the need for and the consequences 
of the confounding alleviation. 

Instead of reparametrizing the random effects vector by decomposing it into two
orthogonal subspaces, we project the neighborhood graph
vertices into the orthogonal space of the covariate matrix $\bm{X}$. 
We transform the geographical centroids inducing a new neighborhood structure that
is not influenced by the covariates. Spatial effects defined on this 
transformed geography will not be correlated with the predictors. 
The projected image of the original graph (hereafter, projected graph)
allows us to keep the sparsity of the precision matrix $\bm{Q}$ and
the Markov properties of random effects, making the estimation of our model 
more efficient than RHZ and HH \citep{rue:book}.
Since SPOCK works only by redefining the non-zero pattern of $\bm{Q}$ by means of the
projected graph, it allows the user to adjust a variety of parameter-based spatial structure
such as \citet{Besag:1974,besag:york:mollie,Leroux,Rodrigues:Assuncao:2012}.

Let $\bm{s} = [\bm{s}_1, \bm{s}_2]$ be the $n \times 2$ matrix with the areas' centroids coordinates.
Assume that $\psi(s_1, s_2)$ is a Gaussian random field conceptually defined for any
location $(s_1, s_2)$ in the map and that  $\bm{\theta} = (\psi_1, \ldots, \psi_n)$
where $\psi_i = \psi(s_{i1}, s_{i2})$. Let $(s_{10}, s_{20})$ be a reference location 
such as a central position in the map. Let $\frac{\partial \psi}{\partial s_{1}}$ and 
$\frac{\partial \psi}{\partial s_{2}}$ be the partial derivatives evaluated at $(s_{10}, s_{20})$. 
When the field is smooth enough, 
the value $\psi(s_1, s_2)$ in an arbitrary location $(s_1, s_2)$ can be written as
\begin{eqnarray}
 \psi(s_1, s_2) &=& \psi(s_{10}, s_{20}) + \left( s_1 - s_{10}, s_2 - s_{20} \right) 
 \left[ \begin{array}{c} \frac{\partial \psi}{\partial s_{1}} \\ \frac{\partial \psi}{\partial s_{2}} \end{array} \right]
 + R(s_1, s_2, s_{10}, s_{20})  \nonumber \\
 &=& \gamma_0 + \gamma_1 (s_1 - s_{10}) + \gamma_2 (s_2 - s_{20})  + R(s_1, s_2, s_{10}, s_{20}) \: .
    \label{eq:randomfield}
\end{eqnarray}
Evaluating the expression \eqref{eq:randomfield} at the centroids $\bm{s}$ and organizing the 
result as a vector, we have
\begin{eqnarray}
\bm{\theta} = \psi(\bm{s}) &=& \gamma_0 \bm{1} + \gamma_1 (\bm{s}_1 - s_{10} \bm{1}) +
\gamma_2 (\bm{s}_2 - s_{20} \bm{1}) + \bm{R}(\bm{s}_1, \bm{s}_2 , s_{10}, s_{20}) \nonumber \\
 &=& (\gamma_0 - s_{10} - s_{20}) \bm{1} + \gamma_1 \bm{s}_1 + \gamma_2 \bm{s}_2 + \bm{R}(\bm{s}_1, \bm{s}_2 , s_{10}, s_{20}) \nonumber \\
 &=&  [ \bm{1}, \bm{s}_1, \bm{s}_2 ] \bm{\gamma} + \bm{R} \: .
    \nonumber
\end{eqnarray}
In this way, we expressed the spatial random effects with a linear component on the coordinates $\bm{s}$.
Returning to the SGLMM model, we can rewrite \eqref{eq:sglmm} as
\begin{eqnarray}
    \bm{Y} &=& \bm{X} \bm{\beta} + [ \bm{1}, \bm{s}_1, \bm{s}_2 ] \bm{\gamma} + \bm{R} \nonumber \\
        &=& \bm{X} \bm{\beta} + \bm{P} [ \bm{1}, \bm{s}_1, \bm{s}_2 ] \bm{\gamma} 
        + \bm{P}^{\bot} [ \bm{1}, \bm{s}_1, \bm{s}_2 ] \bm{\gamma} + \bm{R}  \nonumber \\
        &=& \bm{X} \left( \bm{\beta} + \left( \bm{X}^{\top} \bm{X} \right)^{-1} \bm{X}^{\top} [ \bm{1}, \bm{s}_1, \bm{s}_2 ] \bm{\gamma} \right)   
        + \bm{P}^{\bot} [ \bm{1}, \bm{s}_1, \bm{s}_2 ] \bm{\gamma} + \bm{R}  \nonumber \\
                &=& \bm{X} \bm{\beta}^{*} + \bm{P}^{\bot} [ \bm{1}, \bm{s}_1, \bm{s}_2 ] \bm{\gamma} + \bm{R}  \: ,
        \label{eq:modelconfounded}
\end{eqnarray}
where we used that  $\bm{I} = \bm{P} + \bm{P}^{\bot}$ with $\bm{P}^{\bot}$ defined in 
(\ref{eq:ortogonalP}). 
One of the main advantages of expression \eqref{eq:modelconfounded} is to clearly and intuitively 
answer to \citet{Hodges:Reich:2010}  
when they ask how ``adding spatially correlated errors can mess up the fixed effect you love''. 
The $\bm{X}$ fixed effect is messed up by the spatial random effect $\bm{\theta}$ 
when two conditions are met: (a) the covariates in $\bm{X}$ 
have a linear association with $\bm{s}$ so $\bm{P} [ \bm{1}, \bm{s}_1, \bm{s}_2 ]$ is not 
close to zero and (b) $\bm{\gamma}$ is not small. 
Under these two conditions, the difference between $\bm{\beta}$ and $\bm{\beta}^{*}$ will be large.
If any of these two conditions is not satisfied,
there will be no spatial confounding in the SGLMM regression. This motivates our diagnostic 
tool proposed in Section \ref{s:cca} to verify the need to correct for spatial confounding.

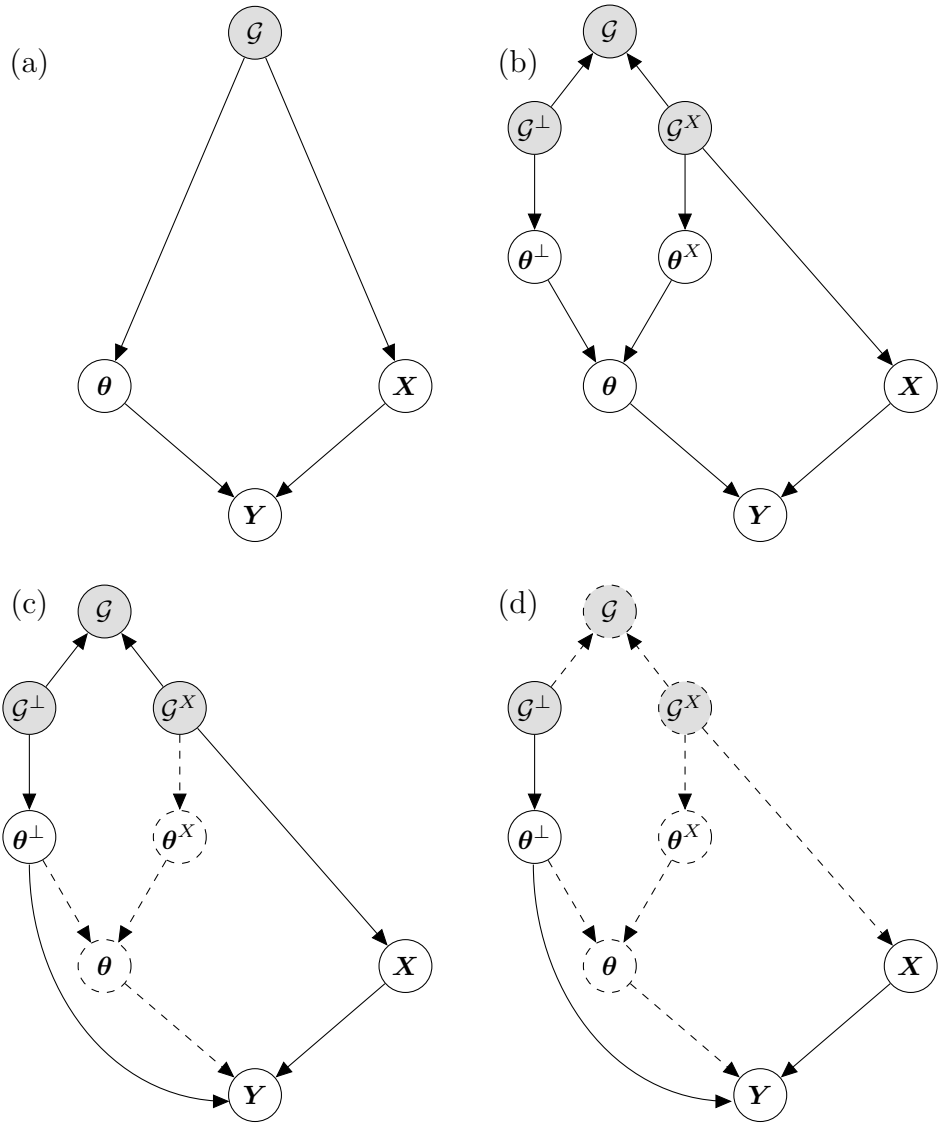
\begin{figure}[!ht]
\begin{tikzpicture}
% \draw[help lines] (-4,7) grid (7,-9);
\tikzstyle{main}=[circle, minimum size = 10mm, thick, draw =black!80, node distance = 16mm]
\tikzstyle{connect}=[-latex, thick]
\tikzstyle{box}=[rectangle, draw=black!100]

% First graph

    % nodes
     \node at (-3,6) {(a)};
     \node[latent] (Y) {$\bm{Y}$};%
     \node[latent,above=of Y,xshift=-2cm] (theta) {$\bm{\theta}$}; %
     \node[latent,above=of Y,xshift=2cm] (X) {$\bm{X}$}; %
     \node[obs, above=of Y, yshift = 4.7cm] (G) {$\mathcal{G}$};%     
    % plate
%     \plate [inner sep=.3cm] {plate1} {(G)} {}; %
    % edges
     \edge {theta} {Y};
     \edge {X} {Y};
     \edge {G} {theta};  
     \edge {G} {X};       

% Second graph 

    % nodes
     \node at (3.5,6) {(b)};
     \node[latent, right=of Y, xshift=5cm] (Y2) {$\bm{Y}$};%
     \node[latent, above=of Y2,xshift=-2cm] (theta2) {$\bm{\theta}$}; %
     \node[latent, above=of Y2,xshift=2cm] (X2) {$\bm{X}$}; %
     \node[latent, above=of theta2, xshift=-1cm] (thetap2) {$\bm{\theta}^{\bot}$};
     \node[latent, above=of theta2, xshift=1cm] (thetaX2) {$\bm{\theta}^{X}$};

     \node[obs, above= of theta2, yshift = 3cm] (G2) {$\mathcal{G}$};%     
     \node[obs, above= of thetap2] (Gp2) {$\mathcal{G}^{\bot}$};%  
     \node[obs, above= of thetaX2] (GX2) {$\mathcal{G}^{X}$};%    

    % edges
     \edge {theta2} {Y2};
     \edge {X2} {Y2};
     \edge {thetap2} {theta2};     
     \edge {thetaX2} {theta2};  
     
     \edge {Gp2} {thetap2};  
     \edge {GX2} {thetaX2};     
     \edge {GX2} {X2};     

     \edge {Gp2} {G2};  
     \edge {GX2} {G2};  
          
% Third graph: RHZ e HH model

    % nodes
     \node at (-3,-1.2) {(c)};
     \node[latent, below= of Y, yshift=-6cm] (Y3) {$\bm{Y}$};%
     \node[latent, dashed, above=of Y3,xshift=-2cm] (theta3) {$\bm{\theta}$}; %
     \node[latent, above=of Y3,xshift=2cm] (X3) {$\bm{X}$}; %
     \node[latent, above=of theta3, xshift=-1cm] (thetap3) {$\bm{\theta}^{\bot}$};
     \node[latent, dashed, above=of theta3, xshift=1cm] (thetaX3) {$\bm{\theta}^{X}$};

     \node[obs, above= of theta3, yshift = 3cm] (G3) {$\mathcal{G}$};%     
     \node[obs, above= of thetap3] (Gp3) {$\mathcal{G}^{\bot}$};%  
     \node[obs, above= of thetaX3] (GX3) {$\mathcal{G}^{X}$};%    

    % edges
     \edge[dashed] {theta3} {Y3};
     \edge {X3} {Y3};
     \edge[dashed] {thetap3} {theta3};     
     \edge[dashed] {thetaX3} {theta3};  
     \draw[->] (-3,-4.65) to [out=270, in=180] (-0.34,-7.8);

     \edge {Gp3} {thetap3};  
     \edge[dashed] {GX3} {thetaX3};     
     \edge {GX3} {X3};     

     \edge {Gp3} {G3};  
     \edge {GX3} {G3};  
          
% Fourth graph: Our model

    % nodes
     \node at (3.5,-1.2) {(d)};
     \node[latent, right=of Y3, xshift=5cm] (Y4) {$\bm{Y}$};%
     \node[latent, dashed, above=of Y4, xshift=-2cm] (theta4) {$\bm{\theta}$}; %
     \node[latent, above=of Y4,xshift=2cm] (X4) {$\bm{X}$}; %
     \node[latent, above=of theta4, xshift=-1cm] (thetap4) {$\bm{\theta}^{\bot}$};
     \node[latent, dashed, above=of theta4, xshift=1cm] (thetaX4) {$\bm{\theta}^{X}$};

     \node[obs, dashed, above= of theta4, yshift = 3cm] (G4) {$\mathcal{G}$};%     
     \node[obs, above= of thetap4] (Gp4) {$\mathcal{G}^{\bot}$};%  
     \node[obs, dashed, above= of thetaX4] (GX4) {$\mathcal{G}^{X}$};%    

    % edges
     \edge[dashed] {theta4} {Y4};
     \edge {X4} {Y4};
     \edge[dashed] {thetap4} {theta4};     
     \edge[dashed] {thetaX4} {theta4};  
     \draw[->] (3.7,-4.65) to [out=270, in=180] (6.34,-7.8);

     \edge {Gp4} {thetap4};  
     \edge[dashed] {GX4} {thetaX4};     
     \edge[dashed] {GX4} {X4};     

     \edge[dashed] {Gp4} {G4};  
     \edge[dashed] {GX4} {G4};  
          
\end{tikzpicture}
\caption{Graphical model representation of the models: (a) usual ICAR model; (b) ICAR with some nodes decomposed; (c) RHZ and HH models; (d) SPOCK model.}
\label{fig:graphicalmodels}
\end{figure}

Expression \eqref{eq:modelconfounded} also justifies SPOCK as a method to deal with spatial 
confounding in spatial regression. Similar to RHZ, we split the linear component of the spatial 
random effect $\bm{\theta}$ 
into two pieces and remove the component $\bm{P} [ \bm{1}, \bm{s}_1, \bm{s}_2 ] \bm{\gamma}$,
which can be confounded with $\bm{X}$. In Figure \ref{fig:graphicalmodels}, 
we show a graphical model representation that explains 
SPOCK and how it differs from the RHZ and HH approach. In $(a)$, we have the usual ICAR model.
The node $\mathcal{G}$ represents the neighborhood graph, which affects both, the covariates $\bm{X}$
and the spatial effects $\bm{\theta}$. The node $\bm{Y}$ is formed according to equation 
(\ref{eq:sglmm}). In $(b)$, we repeat the model in $(a)$ but we decompose some of the nodes 
to contrast the two approaches. 
The node $\bm{\theta}$ is split into two pieces, $\bm{\theta}^{X}$ and  $\bm{\theta}^{\perp}$, following 
the RHZ solution as in \eqref{eq:thetaRHZ}. We also introduce a conceptual decomposition of $\mathcal{G}$ 
into two pieces. The first one is $\mathcal{G}^{X}$, and it carries the shared information between 
the graph and $\bm{X}$. The second one is $\mathcal{G}^{\perp}$, and it contains the residual 
in $\mathcal{G}$ after extracting $\mathcal{G}^{X}$. We explain in the next paragraph what this means and 
how we carry out this decomposition of the neighborhood graph. The RHZ and HH models 
are represented in $(c)$ and they are obtained by removing the node $\bm{\theta}^{X}$ and keeping only the 
$\bm{\theta}^{\perp}$, as shown in \eqref{eq:rhz}. Our SPOCK approach is represented in $(d)$. Differently 
from RHZ and HH, our geography transformation means the removal of $\mathcal{G}^{X}$ and its 
children. One important additional difference is that $\bm{\theta}^{\perp}$ in our model is not the same 
as that defined by RHZ and HH. In our case, $\mathcal{G}^{\perp}$ directly induces a new $\bm{\theta}^{\perp}$
with a sparse precision matrix $\bm{Q}^{\star}$. 

\begin{figure}[!ht]%[tbp]
  \centering
% %   \renewcommand{\subfigbottomskip}{0pt}\renewcommand{\subfigtopskip}{0pt}\renewcommand{\subfigcapskip}{0pt}\renewcommand{\subfigcapmargin}{0pt}
\  \subfigure[]{\label{fig:orig_slov}\includegraphics[width=0.45\textwidth]{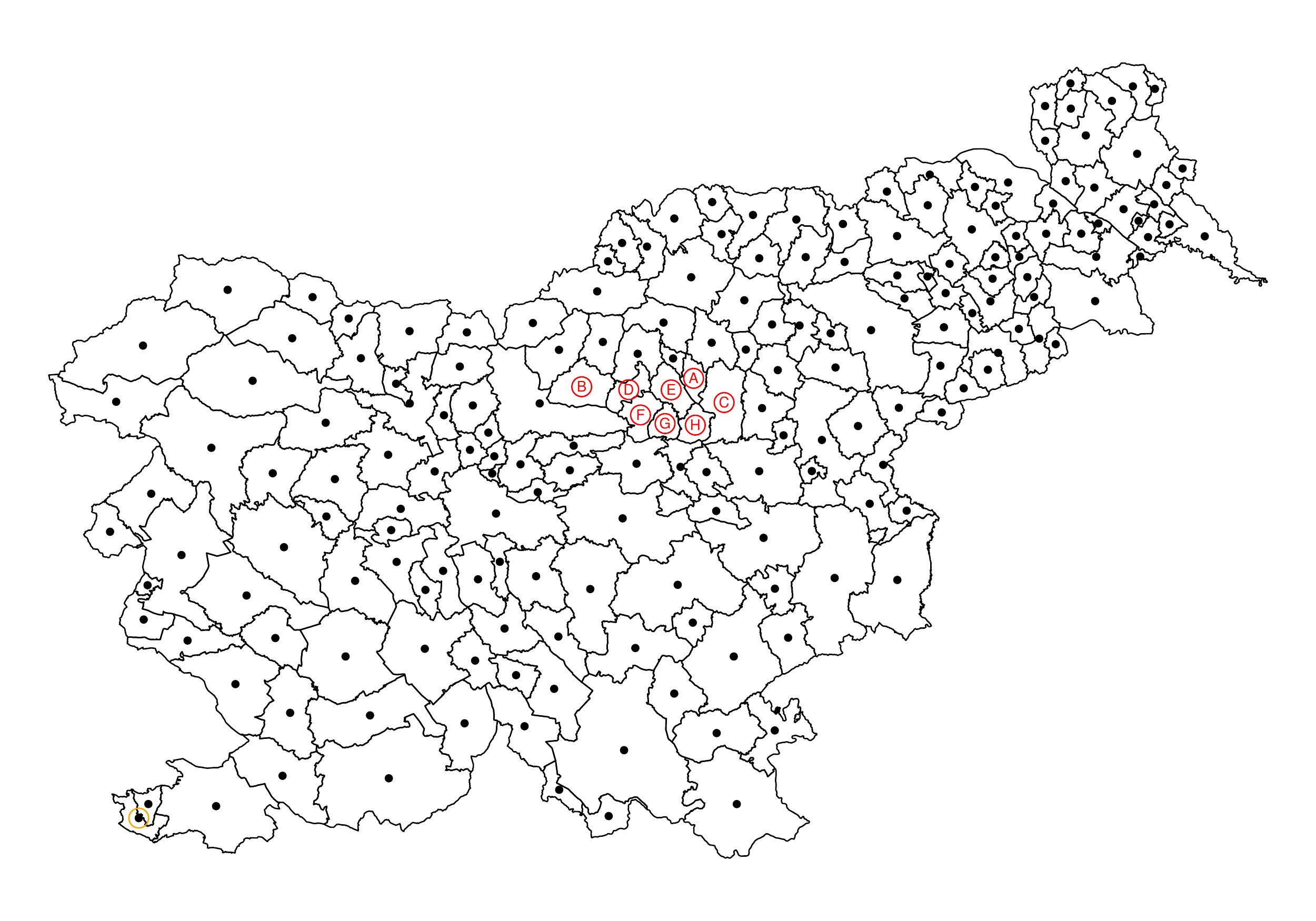}}
  \subfigure[]{\label{fig:proj_slov}\includegraphics[width=0.45\textwidth,height=.25\textheight]{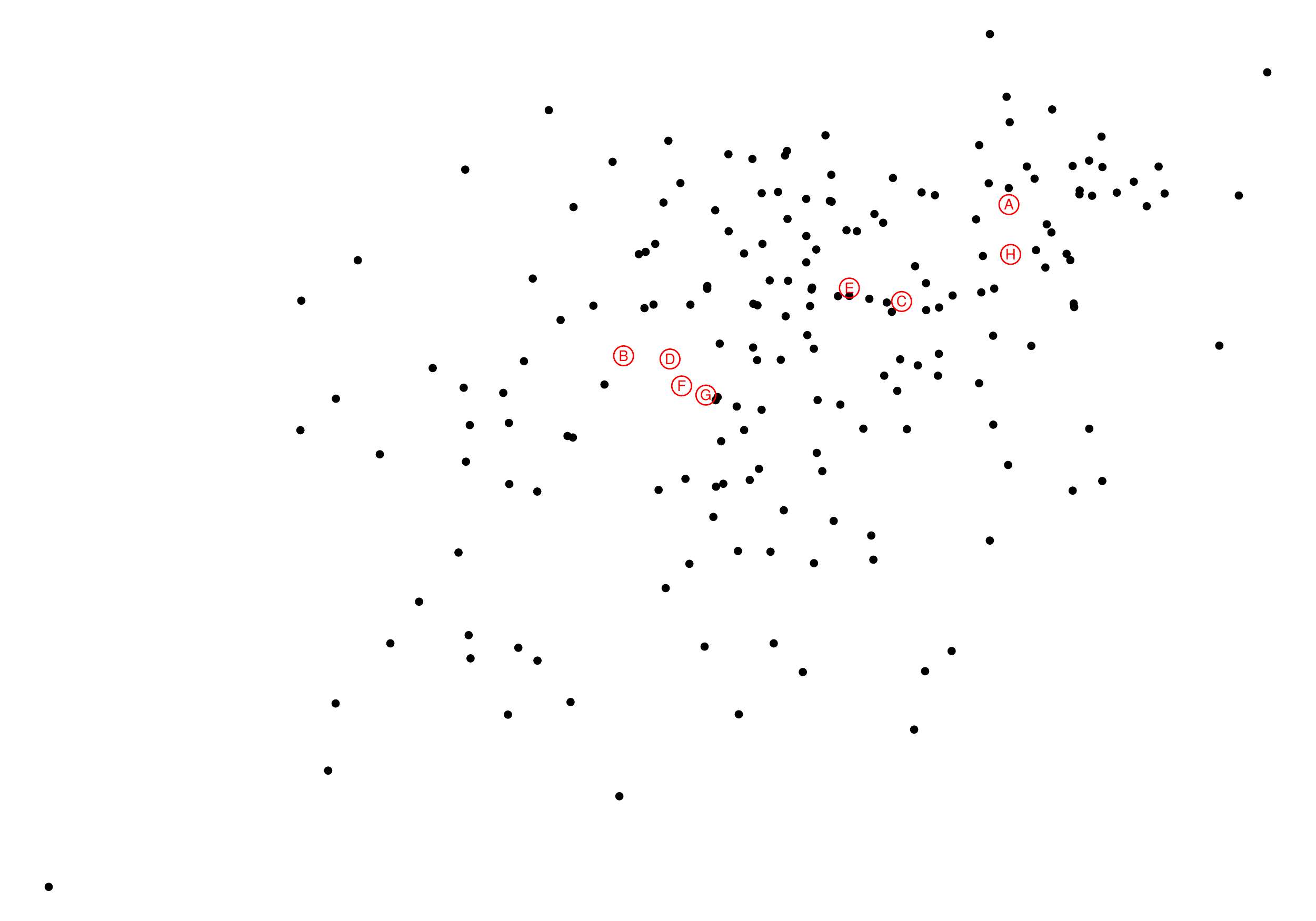}}
  \caption{(a) Original Slovenia map with their areas' centroids. (b) New centroids after projecting the original centroids coordinates into the vector space orthogonal to the columns of $\bm{X}$.}
  \label{fig:slov_proj}
\end{figure}

To explain how we construct $\mathcal{G}^{\perp}$ we use the the same spatial dataset from Slovenia
that motivated \citet{Reich} in their work. Figure \ref{fig:slov_proj}
shows the Slovenia map divided into small regions and the areas' centroids represented by a dot 
inside each polygon. The single predictor variable is a 
social-economic status measure (SES). This predictor has a strong spatial pattern, with a gradient crossing
from the SouthWest to the NorthEast in the map. Our new geography is given by the projected centroids 
$\bm{s}^{\star} = \bm{P}^{\bot} \bm{s}$ shown in the right hand side of Figure \ref{fig:slov_proj}. 
These new set of coordinates are the vertices of the projected graph $\mathcal{G}^{\perp}$ in Figure~\ref{fig:graphicalmodels}. 
In the original map in the left hand side, we selected some areas with red circles to show where they are 
located in the new projected map. We can see that some neighboring areas in the original map are separated out 
and may become isolated from one another in this new geography. 

To define the edges of the $\mathcal{G}^{\perp}$ neighborhood structure,  
we consider two alternatives: 1) to use the number of neighbors $k_i$ of each area 
in the original graph (Knn); 2) to use the Delaunay
triangulation to automatically define the number of neighbors.
In the first method, we add edges between areas $i$ and $j$ if area $j$ is 
among the $k_i$ nearest neighbors of area $i$ in the new geography. 
In the second method, a Delaunay triangulation is
performed over the new centroids and the vertices connected by the triangles' edges 
are considered neighbors. To choose between these two methods, we analysed their ability 
to reproduce the true and original map adjacency structure. After all, when $\bm{X}$ 
is uncorrelated with $\bm{\theta}$, we expect to see only small changes in the spatial 
relationship between the areas. Table \ref{tab:newneigh} shows the result of using the two methods 
in two sets of maps. The first one is composed by the 48 maps of counties from the US states
but Alaska and Hawaii. The second one is composed by 26 maps of counties (indeed, municipalities)
of Brazilian states. In each state, we calculate the \emph{sensitivity} (or precision) of each method: 
the proportion of neighboring pairs of counties in the original state map that is reproduced 
into either the Delaunay or Knn neighborhood map in the new geography. Table \ref{tab:newneigh} 
shows the average sensitivity over all states in US and Brazil. We also calculate the 
\emph{recall} of each method: the proportion of neighboring pairs of counties in the reconstructed 
adjacency structure that is also a neighboring pair in the original map. The higher the better for both,
sensitivity and recall. It is clear from the table that the Knn method is preferred according to 
this criterion and, from now on, all analysis in this manuscript is performed
using the Knn method.

\begin{table}[!htbp]
% \begin{sidewaystable}
\centering
\caption{Percentage agreement in the zero-one pattern between the original graph precision
matrix and the projected precision matrices.}
% \resizebox{\textwidth}{!}{%
\label{tab:newneigh}
\begin{tabular}{c c c  c c}
  \toprule
  & \multicolumn{2}{c}{Sensitivity} & \multicolumn{2}{c}{Recall} \\ 
  \cmidrule(lr){2-3}  \cmidrule(lr){4-5} 
   & Delaunay & Knn  & Delaunay & Knn\\
  \cmidrule(lr){2-2} \cmidrule(lr){3-3} \cmidrule(lr){4-4} \cmidrule(lr){5-5} 
   U.S. counties   & 91.9	& 96.1 & 80.7 & 85.3  \\
   Brazil counties & 91.9	& 96.7 & 80.5 & 86.6   \\
  \bottomrule
\end{tabular}%}
% \end{sidewaystable}
\end{table}

The new adjacency matrix generated by the Knn method replaces the original adjacency
matrix used in the SGLMM model. After that, the user can select his preferred algorithm to fit 
the model and carry out inference. For example, he can use the INLA
\citep{rue:inla} algorithm in order to take advantages of the Markov properties of
the new matrix $\bm{Q}$ generated by the projected graph and to avoid traditional
MCMC convergence problems. Although, we use INLA to perform our analysis in this paper, 
this is not a restriction and other software such as WinBUGS \citep{bugs}, spBayes \citep{Finley:etal:2007}, 
OpenBUGS \citep{Lunn:etal:2009}, or CarBayes \citep{Lee:2013} can be used to adjust SPOCK with the
ICAR or any other prior model for the spatial random effects available.

%We use the Slovenia map (further used in Section~\ref{s:real}) as our baseline map
%for analysis and simulation. A similar example performed in the lattice is performed
%in the real map of Slovenia to see the impacts of the projected graph in a real map.
%In this example we use a explanatory variable without and with spatial effect to see
%how is the project graph.
%In Figure~\ref{fig:slov_rand},

%To better visualize the impact of different spatial trends in the projected graph
%we present a $15 \times 15$ regular lattice example in Figure~\ref{fig:latt1} where %the central area centroid 
%is the origin $(0,0)$. Except by the borders, each area has four neighbors located at one step away at the
%four cardinal directions, as represented by the edges in Figure~\ref{fig:latt1}. Therefore, for most areas, we have $k=4$ neighbors in the Knn method for the projected centroids.

\begin{figure}[!ht]%[tbp]
  \centering
% %   \renewcommand{\subfigbottomskip}{0pt}\renewcommand{\subfigtopskip}{0pt}\renewcommand{\subfigcapskip}{0pt}\renewcommand{\subfigcapmargin}{0pt}
\  \subfigure[]{\label{fig:latt1}\includegraphics[width=0.45\textwidth]{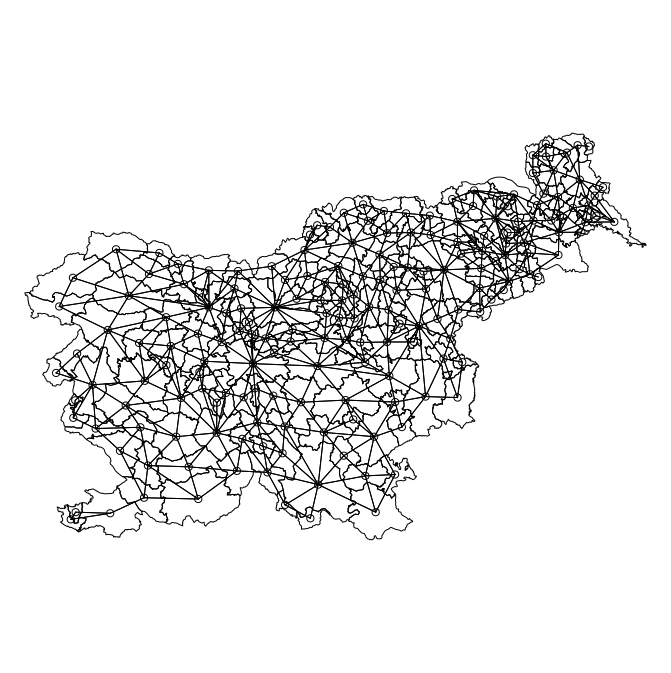}}
  \subfigure[]{\label{fig:lattLN}\includegraphics[width=0.45\textwidth]{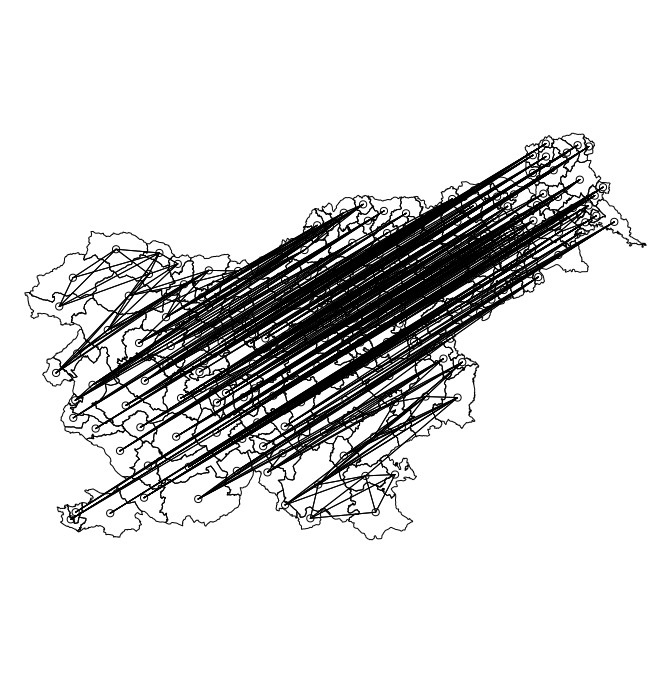}}\\
  \subfigure[]{\label{fig:lattQU}\includegraphics[width=0.45\textwidth]{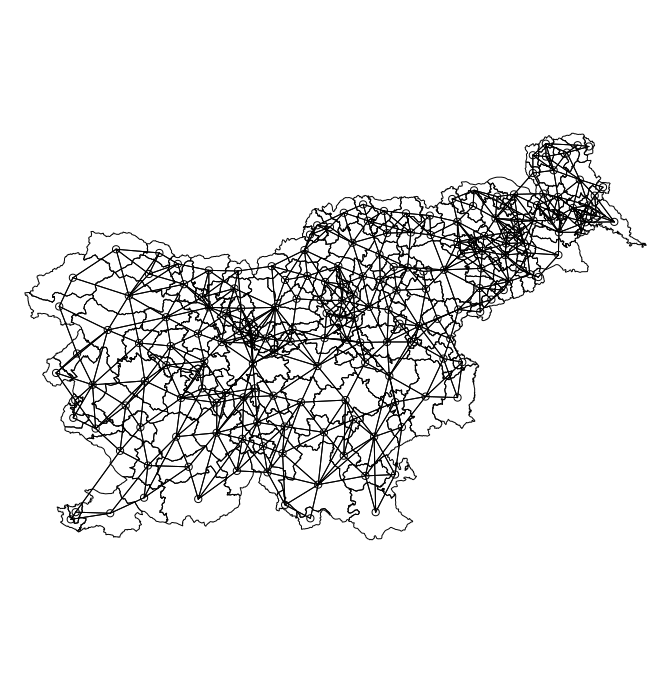}}
  \subfigure[]{\label{fig:lattCU}\includegraphics[width=0.45\textwidth]{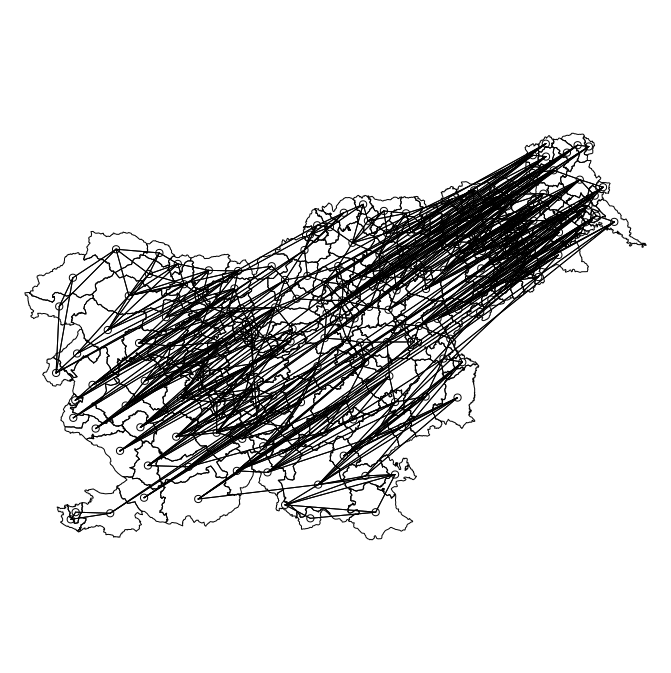}}
  \caption{(a) Original Slovenia map. (b) Reconstructed adjacency matrix
  when $\bm{X}$ is a linear trend of the centroids coordinates. (c) Reconstructed adjacency matrix
  when $\bm{X}$ is a quadratic trend of the centroids coordinates (d) Reconstructed adjacency matrix
  when $\bm{X}$ is a cubic trend of the centroids coordinates.}
  \label{fig:latt}
\end{figure}

To better visualize the impact of different spatial trends in the projected graph
we revisit the Slovenia map example in Figure~\ref{fig:latt1} where the map is now centered at the origin $(0,0)$ and presents its original neighboring structure. For all areas, we keep the $k_i$ number of neighbors in the original map to reconstruct the Knn neighboring structure of the projected geography.
Different spatial trends are created for one explanatory variable $\bm{X}$. 
Figure~\ref{fig:lattLN} is associated with a single covariate given by $X_i = s_{1i}  + s_{2i}$, 
a linear trend based on the centroids' coordinates. Figure~\ref{fig:lattQU} has $X_i = s_{1i}^2  + s_{2i}^2$, 
while Figure~\ref{fig:lattCU} has $X_i = s_{1i}^3  + s_{2i}^3$.

The original centroids are projected into the space generated by $\bm{P}^\bot$. We do not show this new 
geography but rather, in Figure~\ref{fig:latt}, we show the original map with each $i$-th centroid connected to its 
new $k_i$ neighbors in the projected space. Although there is some overlap between segments connecting the new  
neighbors, we can see that each area in Figure~\ref{fig:lattLN} has all its $k_i$ 
neighbors approximately in the $(1,1)$ direction. As $X_i = (1,1) (s_{i1}, s_{i2})^{\top}$, it explains the $Y$ variation 
in the $(1,1)$ direction and so, for the spatial effect, it remains to explain the $Y$ variation in the 
orthogonal $(-1,1)$ direction. In the new geography, far away areas (and therefore, with practically independent 
spatial effects) are those distant in the $(-1,1)$ direction. Hence, the new neighborhood structure must be the 
$k_i$ nearest neighbors in the $(1,1)$ direction. 
When $\bm{X}$ has a quadratic spatial trend, Figure~\ref{fig:lattQU},
there is almost no effect on the projected centroids. This happens because $\bm{X}$ has no linear association with $\bm{s}$ and thus no confounding correction
is necessary. We do not present in the figure, but this pattern is very similar to that 
when $\bm{X}$ is randomly generated (without any spatial pattern).
Finally, Figure~\ref{fig:lattCU} 
presents a cubic pattern for $\bm{X}$. In this case, although not perfect as in Figure~\ref{fig:lattLN},
there is some linear association between $\bm{X}$ and $\bm{s}$ and therefore some correction is needed. 
From Figure~\ref{fig:latt}, it is clear that different spatial trends
generates different effects over the projected graph. This is expected since the
projected graph should be orthogonal to the information in the span of $\bm{X}$.

\section{When do we need to correct for spatial confounding?}
\label{s:cca}

An additional advantage of the SPOCK approach is to provide 
a diagnostic tool for detecting the need for spatial confounding correction.
Statisticians would prefer to use the traditional SGLMM rather than the set of new methods to deal with spatial 
confounding, unless it is really necessary. 
As we saw in Figure \ref{fig:latt}, when $\bm{X}$ has no linear association with the spatial coordinates,
our methodology has no effect on the geography. This could lead one to think that we should always correct for 
spatial confounding as an insurance policy. However, there is a clear price in using the methods that correct
for spatial confounding. They are more complex and harder to interpret by epidemiologists and 
public health agents \citep{Hanks:2015}. 
Next, we provide a tool based on canonical correlation to decide either one 
should do the confounding correction or not. 

To verify the degree of linear association between the covariates $\bm{X}$ and the centroids $\bm{s}$, we apply the canonical correlation technique \citep{johnson1992applied} which measures how much variability the two sets share and if they have some common linear dimensions.  
%Therefore, if $\bm{X}$  and $\bm{s}$  represent similar spatial patterns, the canonical correlation between 
%them will be high and there will be spatial confounding.

The basic idea of the the technique is to find two linear combinations $\bm{U} = \bm{s} \bm{a}$ and 
 $\bm{V} = \bm{X} \bm{b}$, where $\bm{a}$ and $\bm{b}$ are $2 \times 1$ and $q \times 1$, respectively,
 such that the correlation between $\bm{U}$ and $\bm{V}$ is maximized. The 
solution leads to a maximum correlation given by $\rho$, the largest eigenvalue of the 
matrix $\bm{S}^{-1/2}_{ss} \bm{S}_{sx} \bm{S}^{-1}_{xx} \bm{S}_{xs} \bm{S}^{-1/2}_{ss}$,
where 
\[ \bm{S} = \left[ \begin{array}{cc} 
  \bm{S}_{ss} & \bm{S}_{sx} \\
  \bm{S}_{xs} & \bm{S}_{xx} 
  \end{array} \right]
\]
is the empirical covariance matrix of $[\bm{s}, \bm{X}]$.

The need for confounding correction is based on a statistical test of the null hypothesis 
that $\rho=0$. There are two possible approaches, one based on an asymptotic test and another based 
on a Monte Carlo test. 

The asymptotic test is based on the Wilks' Lambda statistic $\Lambda$, which asymptotically has a distribution F \citep{wilks1935independence}. This statistic is a multivariate generalization of the coefficient of 
determination. The basic idea is to fit a multivariate regression model where the multivariate response variable 
is $\bm{s}$ and the covariates are composed by $\bm{X}$. The test statistic measures the proportion of the 
$\bm{s}$ variability that can not be explained by the predictors in $\bm{X}$. 
If this value is small, we have evidence that there is a linear relationship between the two data sets. 
Under the normality assumption for the two matrices, $\bm{s}$ and $\bm{X}$, or if we have a large number of observations, 
we have $\Lambda$ following approximately an $F$ distribution. In cases where these assumptions are not valid, 
we carry out a randomization test by randomly permuting the rows of either $\bm{s}$ or 
$\bm{X}$.

\section{Simulation}
\label{s:sim}

A simulation study is performed with two main goals: 1) to compare the results
of the SPOCK methodology with the existing 
RHZ and HH alternative models, 2) to understand and discuss what are the models assumptions, when it is
necessary to correct for spatial confounding, and what we can obtain from it. The following
model is selected to generate the data:
\begin{eqnarray} \nonumber
   Y_i &=& \beta_0 + \beta_1 X_{1i} + \beta_2 X_{2i} + \theta_i + e_i  \\  \nonumber
   e_i &\stackrel{ind}{\sim}& N(0,\tau_e)   \quad\quad i = 1,\ldots,n 
\end{eqnarray}
where $\bm{\theta}$ is the spatial effect, $n$ is the number of regions in the study.
The coefficients values are $\bm{\beta}^\top = (2, 1, -1)$. We
selected 3 simulation scenarios:
\begin{enumerate}
 \item (ICAR spatial-$X$) The spatial effect is generated using the ICAR structure (that is,
 $\bm{\theta} \sim N(0,\tau_{\theta} \bm{Q})$) with $X_{1i}$ generated from independent standard normal
 distribution and $X_{2i}=s_{i1}$ being the first coordinate of $i$-th area centroid;
 \item (RHZ) The spatial random effect is generated by the orthogonal RHZ proposal: 
 $\bm{\theta} \sim N(0,\tau_{\theta} \bm{P}^\bot \bm{Q} (\bm{P}^\bot)^\top)$.
%  , where $P^\bot$ is the is the orthogonal matrix to the column space of $\mathbf{X}$.
 The explanatory variables are the same as in 1;
 \item (ICAR non-spatial-$X$) Its is equal to scenario 1 except that the explanatory variable $X_{2i}$
 is also generated from independent standard normal distribution without any spatial information.
\end{enumerate}
For each scenario, we generated 1000 datasets with the following combination of $\tau_{\theta}$ and $\tau_e$,
the precision of the random effects  and error term, respectively:
\begin{enumerate}
 \item $\tau_e = 0.2$ and $\tau_{\theta} = 1$;
 \item $\tau_e = 1$ and $\tau_{\theta} = 1$;
 \item $\tau_e = 1$ and $\tau_{\theta} = 0.2$,
\end{enumerate}
to study their effects in model estimation and execution time.
For each of the resulting nine scenarios,
the posterior estimates of the following models were recorded: 1) SPOCK, 2) RHZ, 3) HH, 4) ICAR, 5)linear model (LM), as well as their execution time.
The \texttt{R} software \citep{R} was used to fit all proposed models.
For the RHZ model, we used the \texttt{R} script freely available in 
\url{http://www4.stat.ncsu.edu/~reich/Code/}.
The HH model was fitted using the \texttt{R} package \texttt{ngspatial}
\citep[][version 1.02]{ngspatial}. Finally, to fit the LM, ICAR, and 
SPOCK models, the \texttt{R-INLA} package \citep{rue:inla} was used.

We estimate the posterior mean of the fixed effects in each one of the 1000 replications.
Table~\ref{tab:fixed} presents the median and $2.5\%$ and $97.5\%$ percentiles of these estimates.
We can see that, for all scenarios, the SPOCK model provides very similar inference compared to the 
LM, RHZ, and HH models. The ICAR model has a slightly different behavior.
When the true generating model is the ICAR spatial-$X$, the ICAR model apparently outperforms the other models
as $\tau_{\theta}$ decreases. However, this is not so simple, as we explain soon.
In the RHZ scenario, the ICAR model clearly has a large variance associated with 
$\beta_2$, specially when $\tau_{\theta} = 0.2$. This is expected because, as \citet{Reich} show, 
as the ratio $\tau_{\theta}/\tau_{e}$ decreases, the confounding problem becomes severe. 
Finally, in the ICAR non-spatial-$X$ scenario, with no spatial information in the explanatory variables,
all models seem to present a similar behavior, independently of precision differences.

% \begin{table}[tbp]
\begin{sidewaystable}
\centering
\caption{Median summary of the posterior mean estimate from 1000 replicates in the simulation setting. 
The brackets contain the $2.5\%$ and $97.5\%$ posterior percentiles of fixed effects estimates in 1000 replicates.}
\resizebox{\textwidth}{!}{%
\label{tab:fixed}
\begin{tabular}{c c ccccccccc}
  \toprule
   &  &  \multicolumn{9}{c}{Scenario}\\
  \cmidrule(lr){3-11}
   &           & \multicolumn{3}{c}{ICAR spatial-$X$} & \multicolumn{3}{c}{RHZ} & \multicolumn{3}{c}{ICAR non-spatial-$X$} \\
  \cmidrule(lr){3-5}\cmidrule(lr){6-8}\cmidrule(lr){9-11}
   $\tau_e, \tau_{\theta}$ & Model & $\beta_0$ & $\beta_1$ & $\beta_2$ & $\beta_0$ & $\beta_1$ & $\beta_2$ & $\beta_0$ & $\beta_1$ & $\beta_2$ \\
  \midrule
  True & & 2 & 1 & $-$1 & 2 & 1 & $-$1 & 2 & 1 & $-$1 \\
  \midrule
	  & SPOCK  			    & 2.00 (1.69, 2.31) & 1.00 (0.66, 1.32) & -0.99 (-1.80, -0.12)
					    & 2.00 (1.69, 2.31) & 1.00 (0.67, 1.31) & -1.00 (-1.30, -0.67)
                                            & 2.00 (1.69, 2.31) & 1.00 (0.65, 1.32) & -1.00 (-1.29, -0.62)\\
	  & RHZ				    & 2.00 (1.69, 2.31) & 1.00 (0.66, 1.32) & -0.99 (-1.80, -0.11)
                                            & 2.00 (1.69, 2.31) & 1.00 (0.67, 1.31) & -1.00 (-1.30, -0.67)
                                            & 2.00 (1.69, 2.31) & 1.00 (0.66, 1.32) & -1.00 (-1.29, -0.68)\\
$0.2, 1$  & HH				    & 2.00 (1.69, 2.31) & 1.00 (0.66, 1.32) & -1.02 (-1.84, -0.13)
                                            & 2.00 (1.69, 2.31) & 1.00 (0.67, 1.31) & -1.00 (-1.30, -0.67)
                                            & 2.00 (1.69, 2.31) & 1.00 (0.65, 1.32) & -1.00 (-1.30, -0.67)\\
          & ICAR			    & 2.00 (1.69, 2.31) & 1.00 (0.66, 1.32) & -1.00 (-1.76, -0.18)
                                            & 2.00 (1.69, 2.31) & 1.00 (0.67, 1.30) & -1.00 (-1.46, -0.46)
                                            & 2.00 (1.69, 2.31) & 1.00 (0.66, 1.33) & -1.00 (-1.29, -0.68)\\
	  & LM				    & 2.00 (1.69, 2.31) & 1.00 (0.66, 1.32) & -0.99 (-1.80, -0.12)
                                            & 2.00 (1.69, 2.31) & 1.00 (0.67, 1.31) & -1.00 (-1.30, -0.67)
                                            & 2.00 (1.69, 2.31) & 1.00 (0.65, 1.32) & -1.00 (-1.29, -0.68)\\

  \midrule
	  & SPOCK			    & 2.00 (1.86, 2.14) & 1.00 (0.83, 1.16) & -1.01 (-1.74, -0.18)
                                            & 2.00 (1.86, 2.14) & 1.00 (0.85, 1.14) & -1.00 (-1.14, -0.85)
                                            & 2.00 (1.86, 2.14) & 1.00 (0.83, 1.16) & -1.00 (-1.15, -0.84)\\
	  & RHZ				    & 2.00 (1.86, 2.14) & 1.00 (0.84, 1.16) & -1.01 (-1.75, -0.17)
                                            & 2.00 (1.86, 2.14) & 1.00 (0.85, 1.14) & -1.00 (-1.13, -0.85)
                                            & 2.00 (1.86, 2.14) & 1.00 (0.82, 1.16) & -0.99 (-1.16, -0.83)\\
$1, 1$  & HH				    & 2.00 (1.86, 2.14) & 1.00 (0.84, 1.16) & -1.01 (-1.81, -0.20)
                                            & 2.00 (1.86, 2.14) & 1.00 (0.85, 1.14) & -1.00 (-1.13, -0.85)
                                            & 2.00 (1.86, 2.14) & 1.00 (0.83, 1.16) & -1.00 (-1.16, -0.83)\\
          & ICAR			    & 2.00 (1.86, 2.14) & 1.00 (0.85, 1.16) & -1.01 (-1.63, -0.29)
                                            & 2.00 (1.86, 2.14) & 1.00 (0.84, 1.15) & -1.00 (-1.48, -0.47)
                                            & 2.00 (1.86, 2.14) & 1.00 (0.84, 1.16) & -1.00 (-1.14, -0.85)\\
	  & LM				    & 2.00 (1.86, 2.14) & 1.00 (0.84, 1.16) & -1.01 (-1.75, -0.17)
                                            & 2.00 (1.86, 2.14) & 1.00 (0.85, 1.14) & -1.00 (-1.13, -0.85)
                                            & 2.00 (1.86, 2.14) & 1.00 (0.83, 1.16) & -0.99 (-1.16, -0.83)\\
\midrule
	  & SPOCK			    & 2.00 (1.86, 2.14) & 1.00 (0.76, 1.24) & -1.01 (-2.68,  0.73)
					    & 2.00 (1.86, 2.14) & 1.00 (0.84, 1.14) & -1.00 (-1.14, -0.83)
					    & 2.00 (1.86, 2.14) & 1.00 (0.75, 1.22) & -1.00 (-1.23, -0.76)\\
	  & RHZ				    & 2.00 (1.86, 2.14) & 1.00 (0.76, 1.23) & -0.99 (-2.68,  0.77)
                                            & 2.00 (1.86, 2.14) & 1.00 (0.85, 1.14) & -1.00 (-1.13, -0.85)
                                            & 2.00 (1.86, 2.14) & 1.00 (0.73, 1.24) & -0.99 (-1.26, -0.74)\\
$1, 0.2$  & HH				    & 2.00 (1.86, 2.14) & 1.00 (0.77, 1.23) & -0.99 (-2.79,  0.76)
                                            & 2.00 (1.86, 2.14) & 1.00 (0.85, 1.14) & -1.00 (-1.13, -0.85)
                                            & 2.00 (1.86, 2.14) & 1.00 (0.76, 1.25) & -1.00 (-1.26, -0.73)\\
          & ICAR			    & 2.00 (1.86, 2.14) & 1.00 (0.81, 1.20) & -1.01 (-2.28,  0.36)
                                            & 2.00 (1.86, 2.14) & 1.00 (0.80, 1.19) & -1.03 (-2.20,  0.25)
                                            & 2.00 (1.86, 2.14) & 1.00 (0.80, 1.21) & -1.00 (-1.19, -0.80)\\
	  & LM				    & 2.00 (1.86, 2.14) & 1.00 (0.76, 1.23) & -0.99 (-2.68,  0.77)
					    & 2.00 (1.86, 2.14) & 1.00 (0.85, 1.14) & -1.00 (-1.13, -0.85)
					    & 2.00 (1.86, 2.14) & 1.00 (0.73, 1.24) & -0.99 (-1.26, -0.74)\\
\bottomrule
\end{tabular}}
\end{sidewaystable}
% \end{table}

From Table~\ref{tab:fixed}, we can get two main conclusions. First, there is no major differences in
the estimates provided by the SPOCK and the RHZ or HH models, making it a competitive model. 
The difference on the fitted parameters 
between the ICAR model and the other three models is large when the ratio between the spatial effect precision and
the error precision is small ($\tau_{\theta}/\tau_e = 0.2$). The second one, is the apparent better behavior of the 
ICAR model in the ICAR spatial-$X$ scenario. We discuss that this is not necessarily so in the following.

\citet{Hodges:Reich:2010} discuss the interpretation of confounding in spatial regression and link it with
the multicollinearity problem in linear regression. By construction, the solution proposed by \citet{Reich} and more
recently by \citet{Hughes} assigns to $\bm{X}$ all the spatial variation that $\bm{\theta}$ and $\bm{X}$ are competing for.
That is, RHZ, HH and our own method aims to estimate 
\[ \bm{\beta}^{*} =  \left( \bm{\beta} + \left( \bm{X}^{\top} \bm{X} \right)^{-1} \bm{X}^{\top} 
[ \bm{1}, \bm{s}_1, \bm{s}_2 ] \bm{\gamma} \right)   
\]
in (\ref{eq:modelconfounded}). 
There is not an universal agreement around this solution. 
\citet{Paciorek:2010} and \citet{Hanks:2015} argue that this is a very strong assumption 
to alleviate the confounding problem by making the spatial random
effects orthogonal to the space of the span of $\bm{X}$.

In non-spatial linear models with two covariates $X$ and $Z$
and the presence of multicollinearity, a possible solution arises when one has an 
intrinsic interest on the causal effect of $X$ and $Z$ appears only as nuisance.
In this case, we can regress $Z$ in $X$ and use the regression residual $Z^{\bot}$ as
a covariate instead of $Z$. This situation is exactly what happens in the spatial 
regression problem when you do not want to ``mess up the fixed effect you love'' \citep{Hodges:Reich:2010}. Indeed, in this problem, most of the time, the spatial effect $\bm{\theta}$ has the same role as $Z$ 
and it is only a nuisance error representing 
unobserved spatially structured variables. The covariates in $\bm{X}$ represent truly relevant 
factors with possible causality. This is the rationale to estimate $\bm{\beta}^{*}$ rather than 
$\bm{\beta}$ and it motivates the RHZ, HH, and SPOCK models to alleviate the spatial confounding. 

\begin{figure}[!htb]
  \centering
  \subfigure[]{\label{fig:beta0}\includegraphics[width=0.9\textwidth]{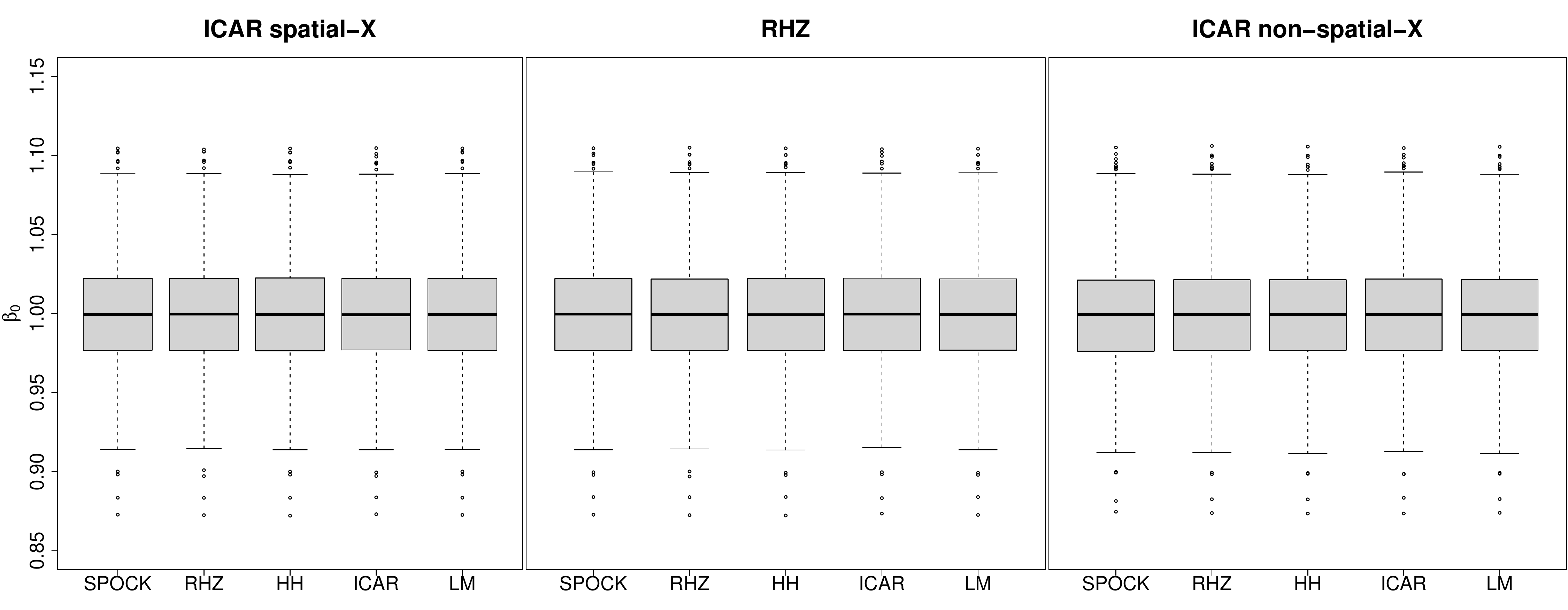}}\\
  \subfigure[]{\label{fig:beta1}\includegraphics[width=0.9\textwidth]{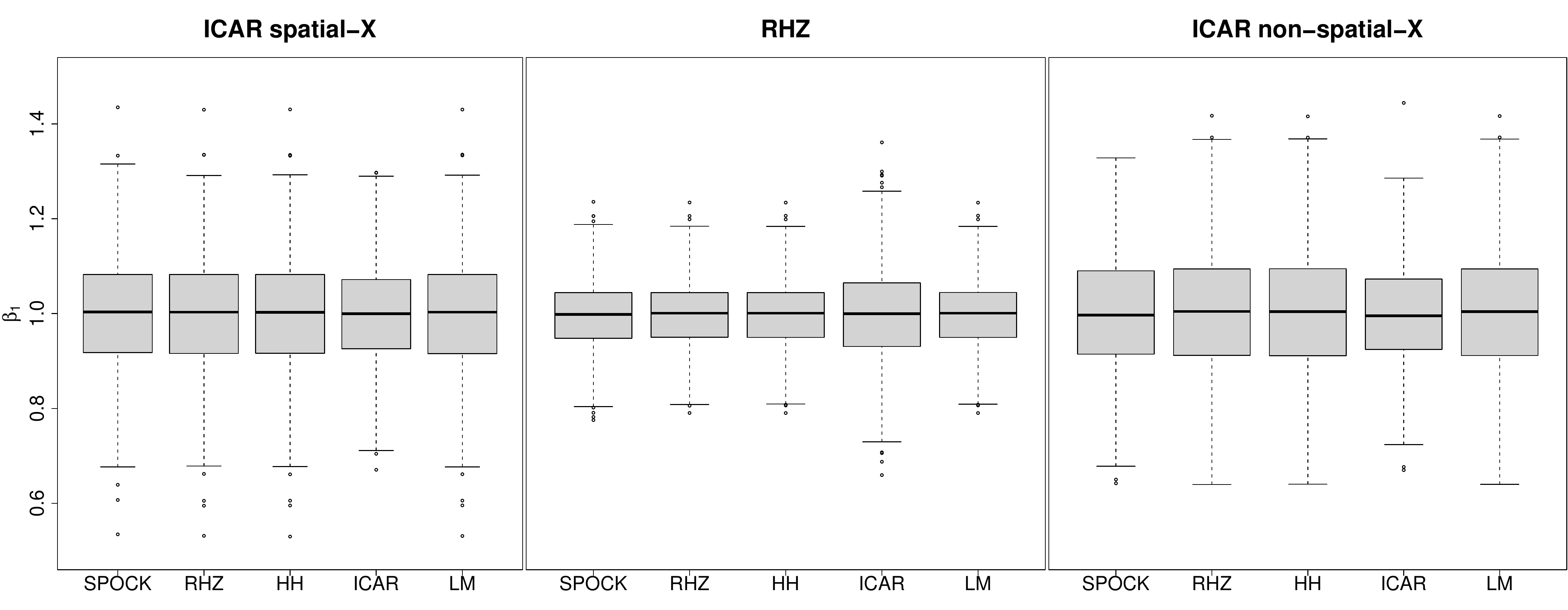}}\\
  \subfigure[]{\label{fig:beta2}\includegraphics[width=0.9\textwidth]{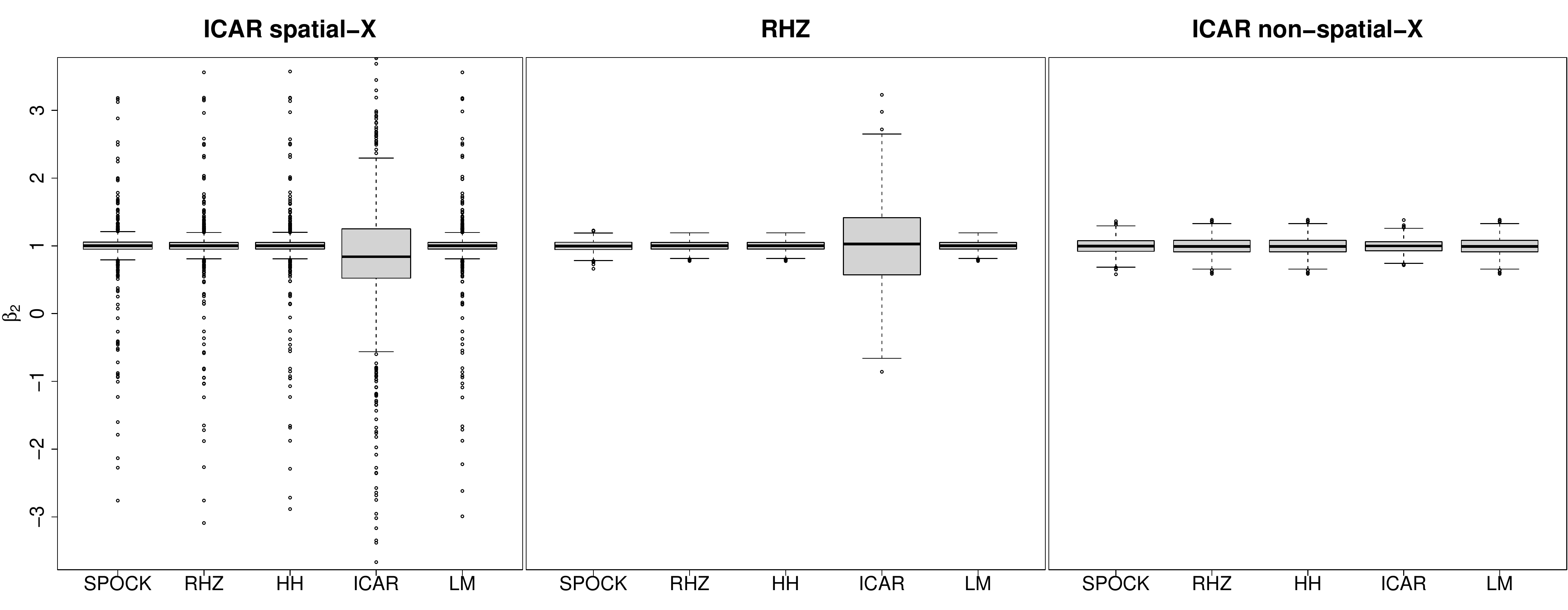}}
  \caption{(a) Posterior estimates for $\beta_0$ under the different models. (b) Posterior estimates for $\beta_1$ under the different
  models. (c) Posterior estimates for $\beta_2$ under the different models.}
  \label{fig:sim_fix}
\end{figure}

A better understanding of the effects of alleviating confounding must be investigated 
looking at the differences between the  $\hat{\bm{\beta}}$ Bayesian estimate (posterior mean) 
and $\bm{\beta}^{*}$, rather than $\bm{\beta}$. To investigate this issue, we produce Figure~\ref{fig:sim_fix}. 
We calculate the ratio $\hat{\bm{\beta}}/\bm{\beta}^{*}$ in each one of the 1000 simulations and show them 
in the box plots for the 3 scenarios with $\tau_{\theta}= 0.2$ and $\tau_e = 1$ fixed. Similar figures 
appear with the other values for $\tau_{\theta}$ and $\tau_e$. 
Each scenario is shown in a different column.  
The first and second rows of plots present $\hat{\beta}_0/\beta^{*}_0$ and $\hat{\beta}_1/\beta^{*}_1$,
respectively. Their coefficients $\beta_0$ and $\beta_1$ are not associated with spatial covariates. 
They have a similar behavior in all scenarios and the main conclusion is that the different methods produce 
very similar results. The third row presents the distribution of the ratio $\hat{\beta}_2/\beta^{*}_2$.
The first two scenarios, ICAR spatial-$X$ and RHZ, have $\beta_2$ associated with a covariate with spatial structure. 
In this case, the traditional ICAR method has a much larger variance than the methods that alleviate spatial confounding.
This shows that the ICAR method is not a good choice when the aim is to assign to $\bm{X}$ all the variation 
in $\bm{\theta}$ that is shared with it. In the ICAR non-spatial-$X$ scenario, all methods return to 
have a similar behavior, as expected. 

After demonstrating that the SPOCK model is capable of removing the spatial confounding and of
providing very similar results to RHZ and HH models, we evaluate the average time
to run each method. Table~\ref{tab:time} presents the median execution time of 1000 replicates
of each model for scenario RHZ. The LM method is substantially faster
than the others. However, the LM method does not take into account the spatial variation that improves
modeling. Among the spatial models, SPOCK clearly outperforms the RHZ and HH methods
in running time and it has comparable time with the ICAR model. The RHZ has the largest median
time, while the HH method seems to perform around 10 times slower than SPOCK. 

\begin{table}[!htb]
% \begin{sidewaystable}
\centering
\caption{Median execution time (in seconds) from 1000 replicates in scenario 2 (RHZ) and the
execution standard deviation (sd) of the evaluation time of each method.}
% \resizebox{\textwidth}{!}{%
\label{tab:time}
\begin{tabular}{c c c c}
  \toprule
   $\tau_e, \tau_{\theta}$ & Model & Time (sec) & sd\\
  \midrule
	  & SPOCK  	& 1.850   & 0.10\\
	  & RHZ		& 157.006 & 0.92\\
$0.2, 1$  & HH		& 19.157 & 3.38\\
	  & ICAR	& 0.597   & 0.09\\
	  & LM		& 0.257   & 0.02\\

  \midrule
	  & SPOCK  	& 1.872   & 0.09\\
	  & RHZ		& 157.152 & 0.62\\
$1, 1$    & HH		& 18.116 & 3.66\\
	  & ICAR	& 0.573   & 0.04\\
	  & LM		& 0.261   & 0.02\\
  \midrule
	  & SPOCK  	& 1.879   & 0.10\\
	  & RHZ		& 158.254 & 1.43\\
$1, 0.2$  & HH		& 20.898 & 3.67\\
          & ICAR	& 0.582   & 0.04\\
	  & LM		& 0.259   & 0.02\\

\bottomrule
\end{tabular}%}
% \end{sidewaystable}
\end{table}

\section{Slovenia Data}
\label{s:real}

The proposed model was adjusted to the same dataset used by \cite{Reich}.
The response variable $Y_i$ is the number of cases of stomach cancer observed in the
municipalities of Slovenia during the 1995-2001 period. The single covariate is the standardized 
value of a socioeconomic status measure  for each area $i=1,\ldots,192$.
Therefore, we have the following model:
\begin{eqnarray} \nonumber
 Y_i|\lambda_i, &\sim& \mbox{Poisson}(\lambda) \\ \nonumber
 \log(\lambda_i) &=& \bm{X}_i \bm{\beta} + \theta_i
 %  \label{eq:jm}
\end{eqnarray}
with $\bm{\beta}$ the fixed effect and $\bm{\theta}$ the spatial effect.

\begin{figure}[!ht]
  \centering
  \subfigure[]{\label{fig:sec}\includegraphics[width=0.45\textwidth]{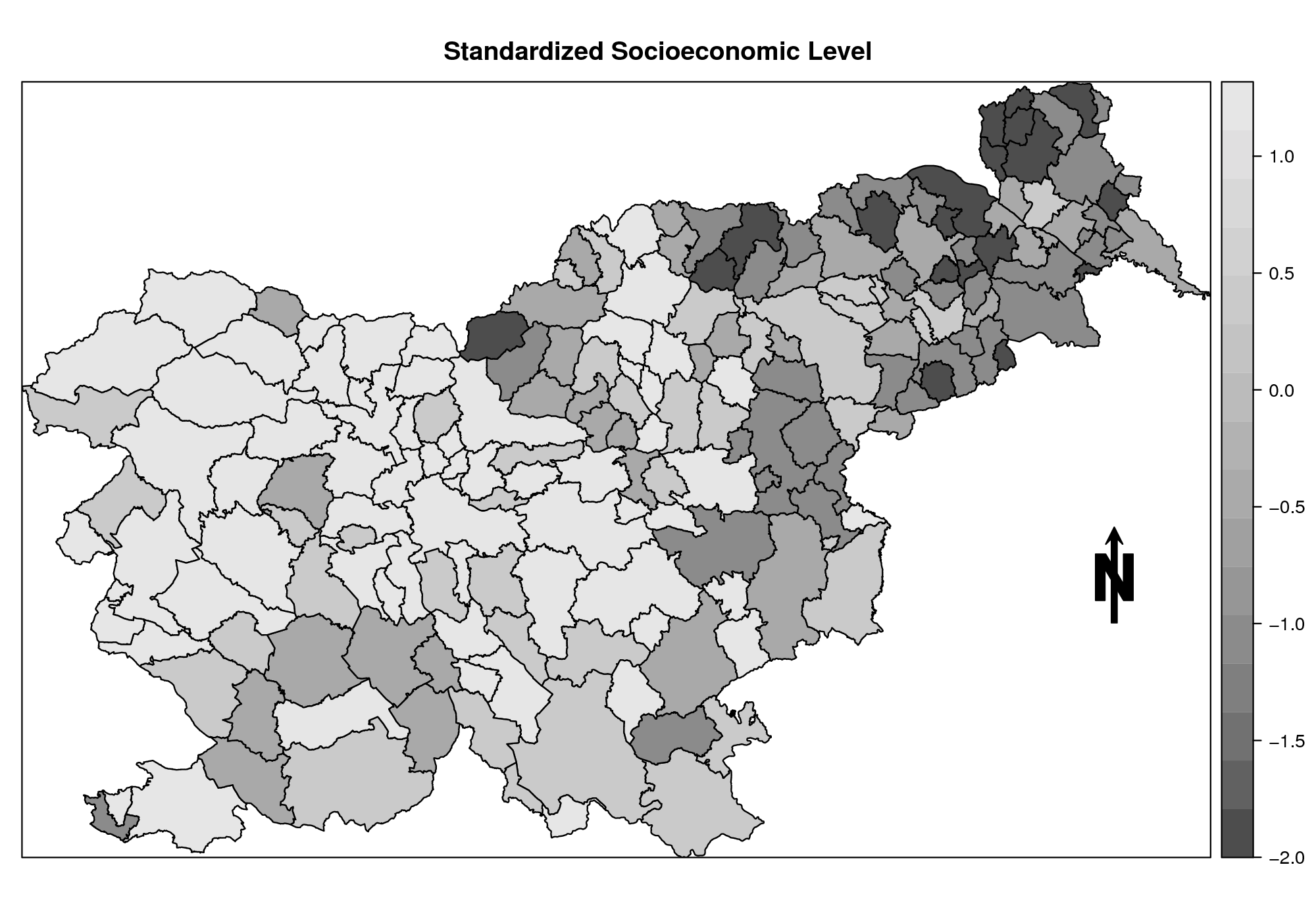}}
  \subfigure[]{\label{fig:icar}\includegraphics[width=0.45\textwidth]{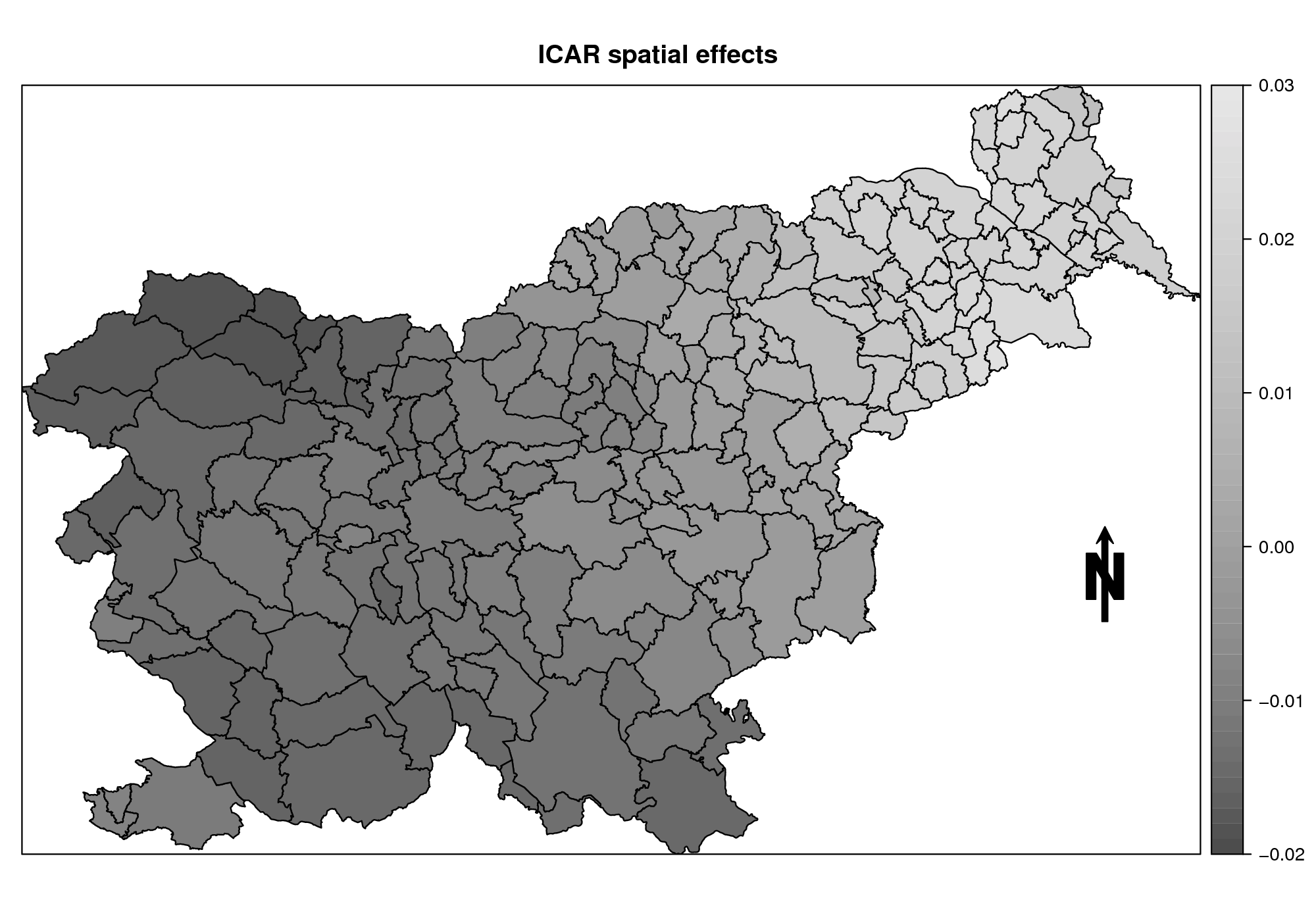}}
  \subfigure[]{\label{fig:ncgmrf}\includegraphics[width=0.45\textwidth]{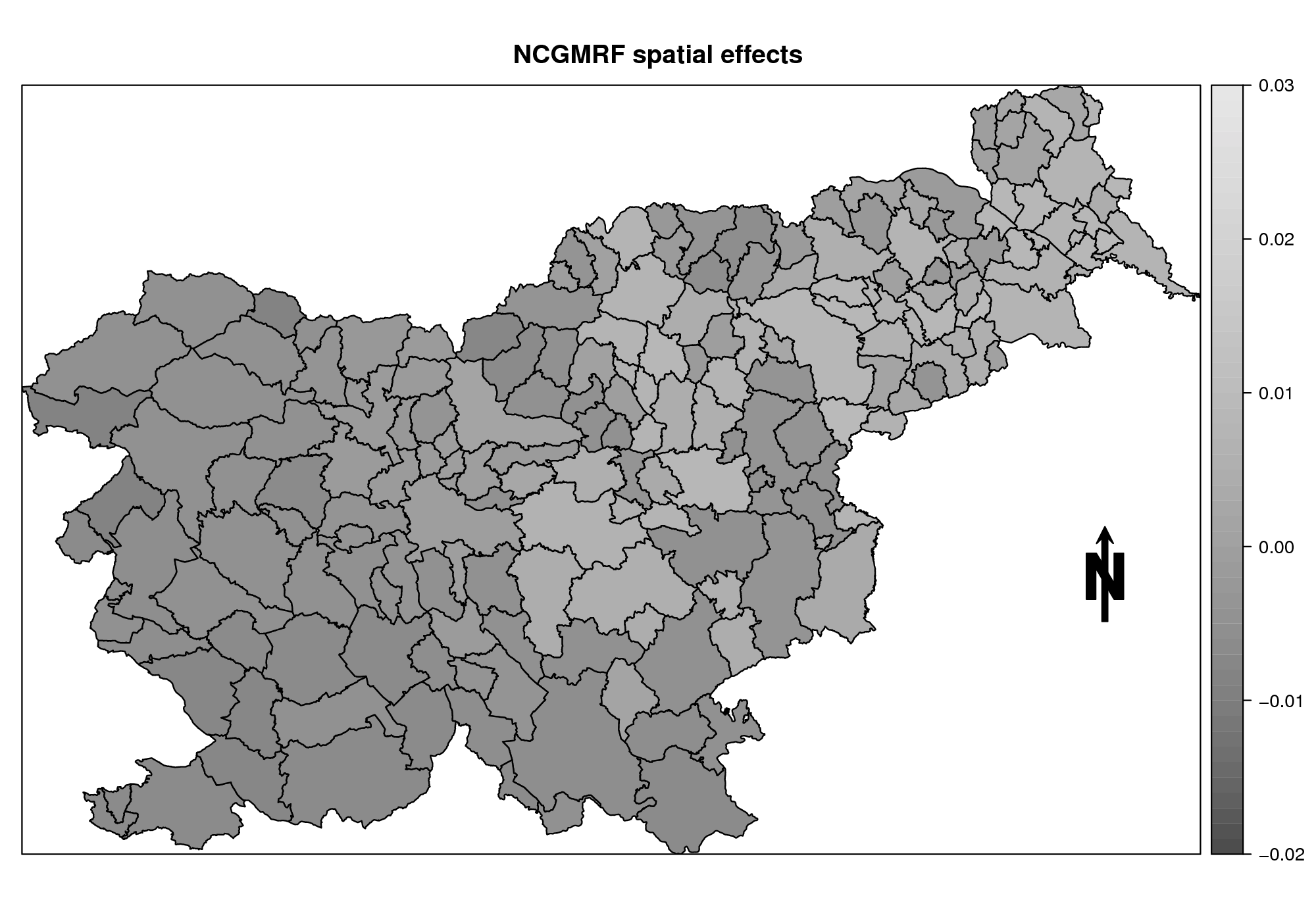}}
  \caption{(a) Standardized socioeconomic level for the municipalities of Slovenia.
  (b) Posterior mean estimates for spatial random effects of the ICAR model.
  (c) Posterior mean estimates for spatial random effects of the SPOCK model}
  \label{fig:slov_sec}
\end{figure}

Using a simple exploratory analysis, the authors noticed that the data show a
negative relationship between the response and the explanatory variable.
This is expected based on the common knowledge of association between health risk and 
deprivation. Their first attempt was to fit the data with the traditional SGLMM
to capture the spatial heterogeneity. However, from Figure~\ref{fig:sec}, it is
clear that the explanatory variable have a diagonal spatial trend presenting higher values
in the southwest and smaller in the northeast. This is an indication of the presence of  
confounding between the random spatial effects and the socioeconomic level.
The pattern in this dataset is similar to what we 
simulated in Section \ref{s:method} when we took $X_i = s_{i1}+s_{i2}$ (see Figure \ref{fig:lattLN}).
After the SGLMM model was fitted by \cite{Reich}, the coefficient
associated with socioeconomic level had a very wide credibility interval that covered even
positive values. That is, the covariate negative effect disappeared. 

Using our diagnosis test from Section \ref{s:cca}, we can verify if there will be, in fact, a confounding effect. Calculating the first canonical correlation between the spatial centroids $\bm{s}$ and the covariate $\bm{X}$, we find 
it equal to 0.67. This value is highly significant according to both, the asymptotic and a permutation test,
obtained from \cite{CCPpackage}. The value is also high in absolute terms, indicating 
that the random spatial effects will be mixed up with the covariate effects.

To better understand the confounding effect between the exploratory
variable and the random effects, we look at the spatial residuals
from the ICAR model. From Figure~\ref{fig:icar},
it is clear that both, the spatial random effect and the explanatory variable, share a
southwest to northeast trend and therefore are competing for the spatial
variability contained in the response $Y_i$.
Figure~\ref{fig:ncgmrf} show the posterior mean estimates for the 
spatial effects under our SPOCK model. After alleviating the 
spatial confounding, there still is some spatial trend left in the same direction 
southwest to northeast. 
However, the spatial dependence now is weaker and the spatial
effects are smoothed toward zero. This is expected since after
alleviating the spatial confounding we assume that the exploratory
variable carries most of the spatial information in $Y_i$. Although weaker,
it is important to notice that the spatial random effect structure
from the SPOCK model is still coherent with the space under analysis.

Table~\ref{tab:cenario3} shows the posterior mean estimates and credible
intervals obtained applying all discussed models. The SPOCK point estimate and
credible interval is similar to the RHZ and HH models.
However, as it can be seen in the last column of Table~\ref{tab:cenario3},
time wise SPOCK drastically outperforms the non-confounding methods.

\begin{table}[htb]
\centering
\begin{tabular}{cccc}
\hline
Model & $\beta_{Se}$& 95\% Credible Interval & Time (sec)\\
\hline
SPOCK & $-$0.1004 & ($-$0.1635, $-$0.0369)  & 2.21 \\
RHZ & $-$0.1137 & ($-$0.1688, $-$0.0596)    & 251.83 \\
HH  & $-$0.1128 & ($-$0.1656, $-$0.0450)    & 29.39 \\
ICAR & $-$0.0474 & ($-$0.1404, 0.0473)      & 0.73 \\
LM  & $-$0.0682 & ($-$0.1067, $-$0.0298)    & 0.16\\
\hline
\end{tabular}
\caption{Posterior mean estimate and credible intervals of
the coefficient associated with the socioeconomic variable (Se) 
for the five fitted models.}
\label{tab:cenario3}
\end{table}

% \begin{table}[h]
% \centering
% \begin{tabular}{c|c|c|c|c}
% \hline
% Model & CAR & RHZ & HH & FNCSGLMM\\
% \hline
% Time (seconds) & 0.53 &519.93&2153.08  &7.81\\
% \hline
% \end{tabular}
% \caption{Processing  time of the four models for data of stomach cancer registered in Slovenia.}
% \label{tab:aplicacao_tempo}
% \end{table}

\section{Conclusion}
\label{s:conc}

In this paper we introduced an alternative way to alleviate spatial confounding.
% named fast non-confounding spatial generalized linear mixed model (NCGMRF).
The main idea is to construct a graph capable of capturing the spatial dependence
orthogonal to the space generated by the span of $\bm{X}$. By doing so, the introduced
method maintains the original sparsity of the precision matrix $\bm{Q}$ and introduces
no restriction in the spatial modeling setup.

From our simulation study and real data example, we were able to show that the SPOCK approach
provides similar results to the others methods that alleviate confounding. These alternative
models project the spatial random effects on the orthogonal space spanned by $\bm{X}$.
Instead, SPOCK projects the original graph $\mathcal{G}$ on that same vector space.
In all simulated scenarios, our method was at least 10 times faster than the existing 
methodologies. This advantage can make it more attractive to researchers in different areas.
Another advantage is that our method can be used with any usual ICAR implementation such as 
INLA, WinBUGS, spBayes, OpenBUGS, and CarBayes. 

We showed that, when no spatial confounding is present, it does not matter which 
method one uses.
However, when this is not the case, running an usual ICAR or the spatial alleviating methods 
may result in different coefficients and inference. The right question is not which approach 
is the correct one. Each of them estimates a different parameter, $\bm{\beta}$ or 
$\bm{\beta}^{*}$. Specific application considerations should guide which parameter is 
more meaningful. In the common situation where the spatial effects $\bm{\theta}$ represent 
geographically structured nuisance effects, it may be desirable to assign to 
$\bm{X}$ all linear spatial variation that is present in $\bm{\theta}$ and hence, to 
estimate $\bm{\beta}^{*}$. However, as mentioned by \citet{Paciorek:2010}, 
the data generating system may have a non-observed spatial
confounder with the observed explanatory variables and setting all spatial variation to the
observed covariates may not be appropriate.

The SPOCK methodology allow for the proposal of a  
diagnostic test that can help on deciding when we should carry out the correction 
for spatial confounding. It is simple and can be calculated previous to any model fitting.
When there is no spatial confounding, fitting ICAR will lead to the same inference as 
the spatial confounding alleviating methods as $\bm{\beta}$ and  
$\bm{\beta}^{*}$ are the same. However, fitting ICAR in the other situation, 
when there is unobserved spatial confounders, leads to an estimate of  
$\bm{\beta}$, rather than of $\bm{\beta}^{*}$. The user should be 
aware of these differences so he does not use one thinking to have the other.

\section*{Acknowledgments}
The authors acknowledge CNPq, CAPES, and FAPEMIG for partial financial support.

\bibliographystyle{asa}
\bibliography{main}

\end{document}